# Structural features, stacking faults, and grain boundaries in $MgB_2$ superconducting materials

J.Q. Li[a], L. Li[a,b], Y.Q. Zhou[a], Z. A. Ren[a], G.C. Che[a], and Z.X. Zhao[a]

[a] National Laboratory for Superconductivity, Institute of Physics, Chinese Academy of Sciences, Beijing 100080, P. R. China
[b] Department of Physics, Ningxia University, Yinchuan, Ningxia 750021, P.R. China.

**The structural properties of $MgB_2$ superconductors have been analyzed by means of convergent-beam electron diffraction, high-resolution transmission-electron microscopy, and theoretical simulations. The $MgB_2$ crystal has been identified to have a hexagonal structure with the space group of P6/mmm. Microstructural features of $MgB_6$ materials depend evidently upon the synthesis pressures. Stacking faults and grain boundaries in a sample with the critical current density greater than $10^5 A/cm^2$ have been extensively investigated.**

## 1. INTRODUCTION

Recently, Akimitsu et al [1] announced a new binary intermetallic superconductor, magnesium diboride ($MgB_2$), with the superconducting critical temperature at around 39K. This discovery stimulated great interest in the investigations of the electronic, magnetic, and structural properties of this kind of materials [2-5]. The superconducting transitions in the $MgB_2$ materials synthesized under a variety of conditions seem to be at around the limit of $T_c$ as suggested theoretically several decades ago for BCS, phonon-mediated superconductivity. The magnetic measurements showed that $MgB_2$ materials prepared under a high pressure of 6Gpa has not only a high $T_c$ but also a large critical current density ($J_c > 10^5 A/cm^2$) under the applied magnetic fields less than 1T [6]. It is known that the microstructural features could essentially affect the critical current of superconducting materials. In this paper, we reported on the investigation of the structural properties of the $MgB_2$ superconducting materials prepared under a variety of conditions. Grain boundaries and planar defects in a sample with the critical current density greater than $10^5 A/cm^2$ are in particular discussed.

## 2. EXPERIMENTAL

A series of $MgB_2$ samples prepared under both the ambient condition and high pressures were used for the present study. A report on the sample preparation, superconductivity, and x-ray analyses has been published in ref. 6. Specimens for transmission-electron microscopy (TEM) observations were polished mechanically with a Gatan polisher to a thickness of around 50μm, dipped to 20μm, and then ion-milled by a Gatan-691 PIPS ion miller for 3 h. The TEM investigations were performed on a H-9000NA electron microscope with an atomic resolution of about 0.19nm.

## 3. RESULTS

TEM observations reveal that the $MgB_2$ crystals in all samples have a well-defined $AlB_2$-type hexagonal structure with the space group of P6/mmm as reported previously in ref. 1. The essential features of this structure can be clearly illustrated with convergent-beam electron diffraction (CBED) patterns and TEM images obtained along several relevant zone-axis directions. Figure 1(a) shows the [001] zone-axis CBED pattern, illustrating the 6-mm symmetry with the systematical mirror-planes on the {100} and {210} crystal planes. A sixfold axis is along the c direction. Figure 1(b) is the [010] zone-axis CBED pattern, illustrating the 2-mm symmetry with mirror symmetries on the (001) and (010) crystal planes and a two-fold rotation axis in the <010> direction. It is known that the radii of the high order Laue-zone rings in the CBED pattern relate to the crystal periodicity along the incident-beam direction. Thus corresponding to the Laue-zone ring in Fig.1 (a), the c parameter of the unit cell is estimated to be about 0.36nm. All diffraction spots shown in these patterns can be well indexed by a hexagonal cell with lattice parameters of a=b=0.31nm, and c=0.352nm. In order to facilitate the comparisons, we show two calculated CBED patterns in Fig. 1(c) and (d) taken along the [001] and [010] zone-axis directions, respectively. All symmetry properties revealed in these theoretical patterns are perfectly consistent with the experimental ones.

Better and clear views of the atomic structure of $MgB_2$ superconductor have been obtained by high-resolution TEM observations. Figure 2 (a) and (b) show the high-resolution electron micrographs of $MgB_2$ crystal taken along the <001> and <010> zone-axis directions, respectively. These images were obtained from thin regions in the $MgB_2$ crystals; therefore, we expect that, in combination with the results of theoretical simulations, the Mg- or B-atom positions could be identified. Image calculations, based on the schematic models as displayed in the insets of the images, were carried out by varying the crystal thickness from 2 to 5 nm and the defocus value from -10 to -60 nm. A calculated image with the defocus value of –0.49nm and the thickness of 2.5nm is superimposed on the Fig. 2(a), and appears to be in good agreement with the experimental one. In this image the Mg-atom positions are recognizable as bright dots. On the other hand, B atoms give rise to a very low contrast in this image. The TEM image in Fig. 2 (b) is likely to be obtained under the defocus value at around the Scherzer defocus ( ~-32nm). The Mg and the B atomic columns are therefore recognizable as dark dots. The layered structural feature, i.e. an alternation of the hexagonal Mg layer and the graphite-like honeycomb B layer along the c direction, can be clearly read out. A calculated image for a defocus of –32nm and a thickness of 2.8nm, superimposed onto Fig. 2(b), and appears to be in good agreement with the experimental one. It is hardly surprising that the experimental images have not yielded the perfectly identifiable contrast at the B-atom positions, owing to the short distance between two adjacent B-atom columns. For instance, the inset of Fig. 2(b) illustrates schematically the projection of $MgB_2$ crystal along [010] zone-axis, the shorter distance between two adjacent B-atom columns is about 0.09nm, which is too short to be resolvable in our electron microscope with a resolution of ~0.19nm. Hence, these two columns are recognizable as one dark dot in the experimental image.

Microstructural features of $MgB_2$ materials depend essentially on the synthesis conditions; especially develop with the increase of pressure. Materials prepared under the ambient pressure contain discernable empty space as well as impurity phases at the grain boundaries. Materials synthesized under a high pressure larger than 3Gpa are evidently denser; the empty spaces among grains almost disappear. Samples prepared under a high pressure of 6Gpa consist of regularly packed large grains ranging from 0.4 to 10 μm in sizes, these materials also give rise to a high Jc (> $10^5 A/cm^2$) as measured experimentally and estimated theoretically following with the Bean critical-state model [6]. Figure 3(a) is a bright-field TEM image showing the microstructure of a material prepared under a high pressure of 3Gpa. The crystalline grains in this sample range from 0.1-0.8μm in sizes, these grains are randomly packed together without any orientation coherence. In figure 3(b) to (d), we show a series of TEM images of a material synthesized under a high pressure of 6Gpa. It is worth noting that the grains become apparently larger with an average size over 1 μm, these grains in many regions appears to be regularly packed either along the c-axis direction or within the a-b crystal plane. Figure 3(c) shows a set of $MgB_2$ grains is fitly stacked along the c-axis direction. They are well connected without impurity phases at the grain boundaries. Figure 3(d) shows a bright-field TEM image illustrating the microstructural properties of $MgB_2$ grains within the a-b plane. The ordered arrangement of the large grains also appears in this area. On the other hand, the impurities, creating evidently different contrast at certain grain boundaries, become visible as typically indicated by arrows. These impurity materials could be MgO or $MgB_4$ as detected by means of energy dispersive x-ray (EDX) microanalysis.

Stacking faults as well as some other kinds of planar defects within the a-b plane frequently appear in the large grains. Figure 4(a) shows a TEM image revealing several stacking faults in a $MgB_2$ grain. These planar defects locate on the a-b crystal plane and result in apparent structural distortions in the vicinal areas. Figure 4(b) displays the corresponding electron diffraction pattern, illustrating the diffuse diffraction spots streaking along the c*-axis direction. Figure 4(c) shows a high-resolution TEM image of the rectangular area in Fig.4 (a), clearly revealing the structural properties of a stacking fault. The structural distortion and lattice mismatch are recognizable nearby this planar defect. The frequent appearance of this kind of stacking faults originates from the weak coupling between the B- and Mg-atomic layers. Structural models for interpreting the arrangements of B and Mg atoms nearby the faults with either an addition atomic layer or some types of dislocations will be reported in a coming paper.

In order to understand the structural properties of grain boundaries with impurity phases, we have performed EDX analyses in combination with nano-diffraction measurements. Figure 5(a) is a bright-field TEM image illustrating a small grain of an impurity phase trapped among three $MgB_2$ grains. Systematical analyses suggest it is a $MgB_4$ grain as documented by the EDX spectra shown in Fig. 5(b). As a matter of fact, impurity materials observed at many grain boundaries are found to $MgB_4$, and usually their sizes ranges from 10 to 300nm. The $MgB_4$ material has an orthorhombic structure with lattice parameters of a=0.54nm, b=0.44nm and c=0.74nm. Furthermore, high-resolution TEM observations have revealed that the $MgB_2$ grains could coherently grow with $MgB_4$ impurities either along the c-axis direction due to lattice fitness of $c(MgB_4) \approx 2c(MgB_2)$ or within the a-b plane due to $a(MgB_4) \approx 2d_{100}(MgB_2)$. Figure 5(c) shows a high-resolution TEM image of the rectangular area between grain a and grain b, illustrating the structural features of a grain boundary with the $MgB_4$ impurity. This $MgB_4$ grain grows coherently with a $MgB_2$ grain within the a-b plane. Figure 5(d) shows a high-resolution TEM image of the boundary between grain a and grain c. These two grains with different orientations are well connected without impurity between.

## 4. CONCLUSION

The structural features of $MgB_2$ superconductors depend systematically upon the conditions of sample preparation, especially, the synthesis pressures. The $MgB_2$ materials prepared under 6Gpa consist of large, regularly packed grains ranging from 0.1 to 8 $\mu m$ in sizes. Stacking faults as well as other kinds of planar defects appear frequently in this kind of superconducting materials. Most grain boundaries are occupied by the $MgB_4$ materials which could grow coherently with the $MgB_2$ superconducting grains.

**Acknowledgments**

The authors would like to express many thanks to Prof. C. Dong, Prof. D. N. Zhen, and Prof. L. P. You for their assistances. The work reported here was supported by "Hundreds of Talents" program organized by the Chinese Academy of Sciences, P. R. China.

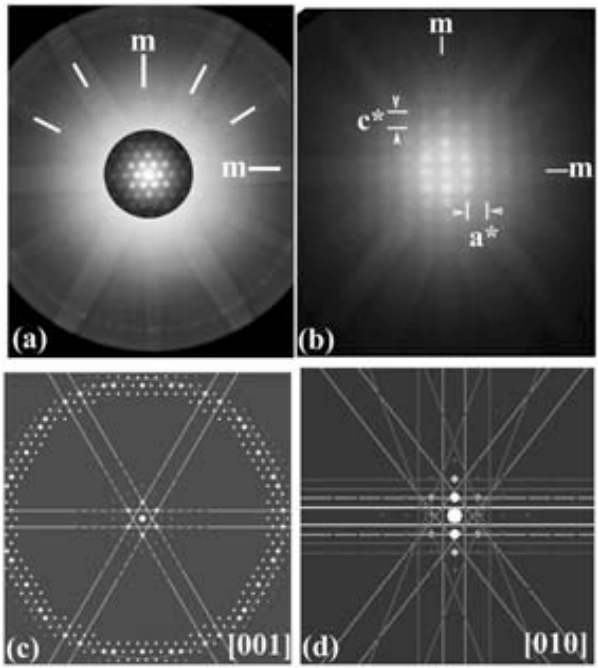

FIG. 1. CBED patterns of the MgB$_2$ superconductor. (a) [001] zone-axis pattern showing 6mm symmetry. (b) [010] zone-axis pattern showing 2mm symmetry. (c), (d) Calculated CBED patterns showing the symmetry features in agreement with the experimental patterns.

FIG. 2. High-resolution TEM images of MgB$_2$ taken along (a) [001] zone-axis direction and (b) zone-axis direction. The insets show schematically the relevant structural models and theoretical images.

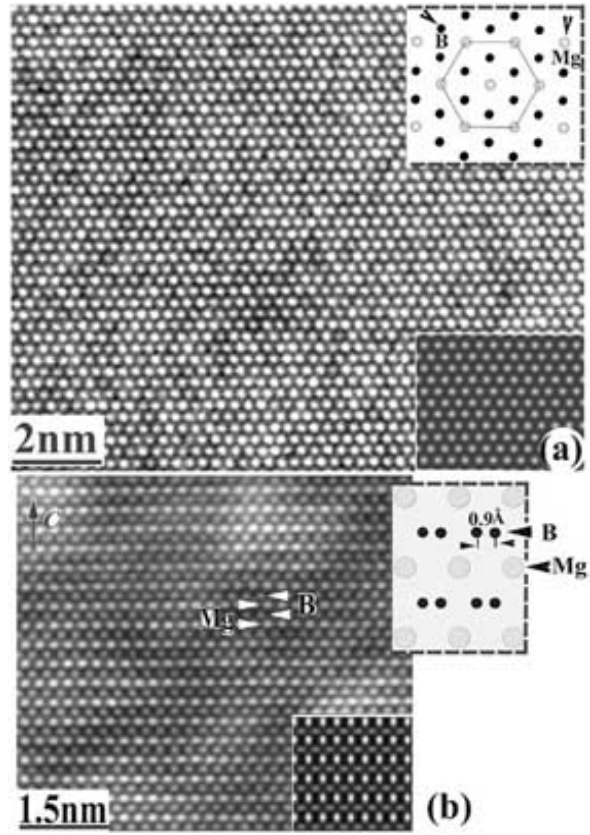

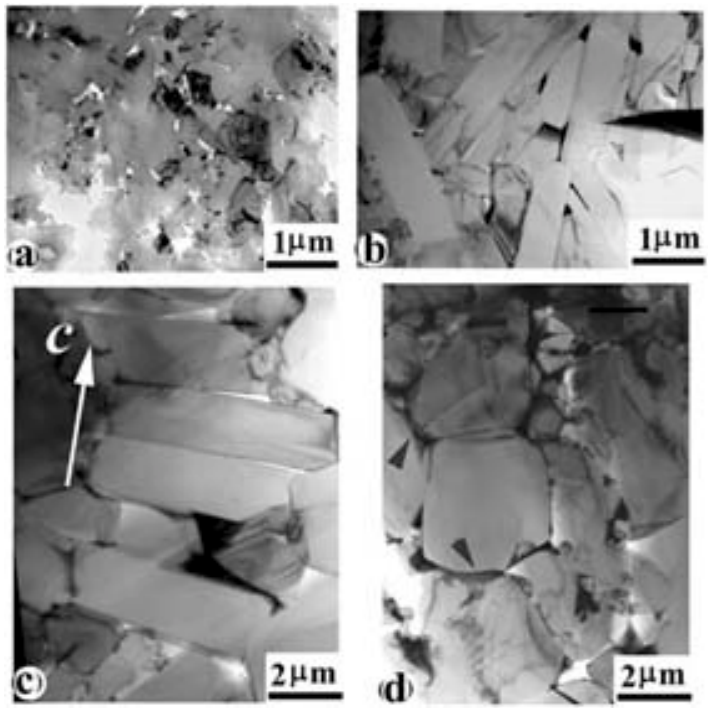

FIG. 3. Bight-field TEM images showing the structural properties of (a) a material prepared under 3Gpa with grain sizes of 0.1 to 0.8 μm; (b) a material prepared under 6Gpa with grain sizes of 0.4-8μm; (c) regularly packed $MgB_2$ grains along c-direction; and (d) large $MgB_2$ grains with impurities at some boundaries.

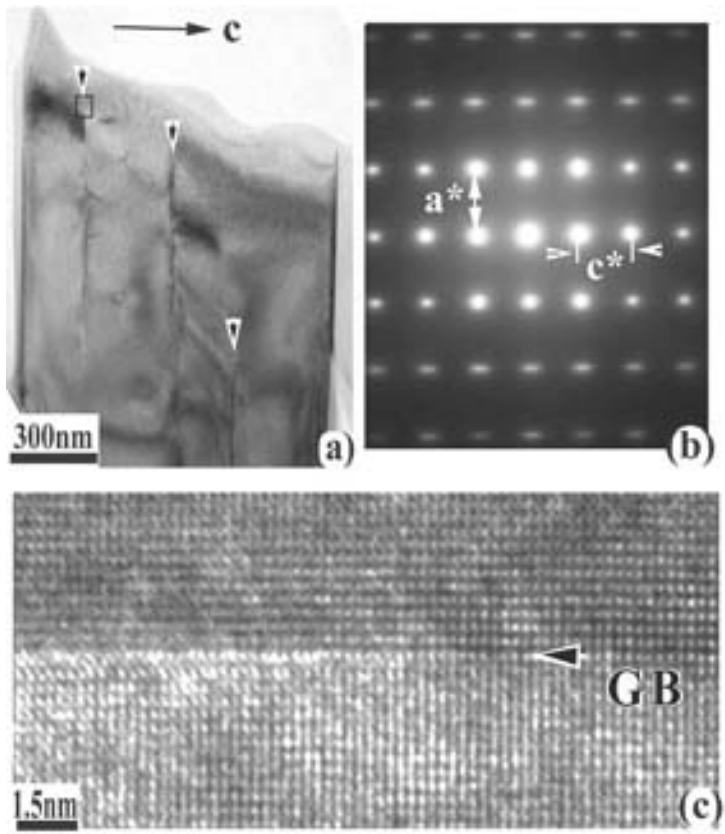

FIG. 4. (a) Bright-field TEM image showing stacking faults in a $MgB_2$ grain. (b) Electron diffraction pattern showing the broad spots arising from stacking faults. (c) High-resolution TEM image showing the structural distortion in the vicinal area of a stacking fault.

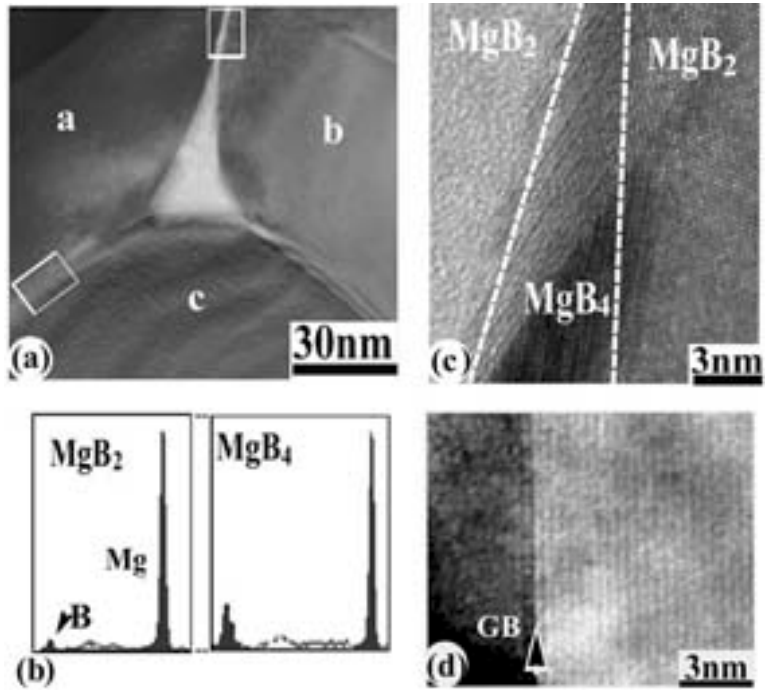

FIG. 5. (a) TEM image of grain boundaries with $MgB_4$ impurity. (b) EDX results from the $MgB_2$ grains (left) and from the impurity material (right). (c) High-resolution TEM image of a grain boundary with $MgB_4$, it grows coherently with a $MgB_2$ grain. (d) Grain boundary between grain a and grain c; no intergranular impurity is visible in this region.